\documentclass[useAMS,usenatbib]{mn2e}
\usepackage{epsfig}
\usepackage{amssymb}
\usepackage{amsmath}
\usepackage{lscape}
\usepackage{longtable}
\usepackage[normal]{caption}
\usepackage{appendix}
\usepackage{bm}

\title[What produces the far-infrared/submm emission in the most luminous QSOs?]{What produces the far-infrared/submm emission in the most luminous QSOs?}

\author[M.~Symeonidis] 
{\parbox{\textwidth}{\raggedright
M.~Symeonidis$^{1}$\thanks{E-mail: \texttt{m.symeonidis@ucl.ac.uk}}
}\vspace{0.4cm}\\
\parbox{\textwidth}{\raggedright $^{1}$ Mullard Space Science
  Laboratory, University College London, Holmbury St. Mary, Dorking,
  Surrey RH5 6NT, UK\\
}}

\begin{document}

\date{Accepted  Received; in original form}

\pagerange{\pageref{firstpage}--\pageref{lastpage}} \pubyear{2014}

\maketitle

\label{firstpage}

\begin{abstract}
The AGN. 
\newline
I examine the average spectral energy distributions (SEDs) of two samples of the most powerful, unobscured QSOs at $2<z<3.5$, with rest frame optical luminosities in the 46.2$<$log\,$\nu L_{\nu}$(\rm 5100\,\AA)$<$47.4 range, corresponding to the tail of the $2<z<4$ QSO optical luminosity function. I find that the AGN could potentially account for the entire broad-band emission from the UV to the submm, on the basis that the SEDs of these sources are similar to the \textit{intrinsic} AGN SEDs derived for lower power, lower redshift QSOs. Although this does not preclude substantial star-formation in their host galaxies, I find that the AGN dominates the total infrared luminosity, removing the necessity for a star-forming component  in the far-IR/submm. I argue that the origin of the far-IR/submm emission in such powerful QSOs includes a small contribution from the AGN torus, but is predominantly linked to dust at kpc-scales heated by the AGN. The latter component accounts for at least 5-10 per cent of the bolometric AGN luminosity and has an implied dust mass of the order of 10$^8$\,M$_{\odot}$.

\end{abstract}


\section{Introduction}
\label{sec:introduction}
It is now well established that the most energetic, non-transient sources in the Universe are AGN, the most optically luminous of which are often referred to as quasi-stellar objects (QSOs). The first QSOs were identified via their radio emission (Greenstein $\&$ Matthews 1963; Schmidt 1963), but in the last few decades the number of known luminous and/or high redshift QSOs has significantly increased thanks to large scale optical surveys such as the SDSS (Alam et al. 2015\nocite{Alam15}) and 2dF/6dF (Colless et al. 2001\nocite{Colless01}; Jones et al. 2009\nocite{Jones09}) and near-IR surveys such as 2MASS (Skrutskie et al. 2006\nocite{Skrutskie06}) and UKIDSS (Lawrence et al. 2007\nocite{Lawrence07}). The last few years saw the discovery of the most luminous QSO, SDSS\,J010013.02+280225.8, found at $z=6.3$ with a bolometric luminosity of 1.62$\times$10$^{48}$\,erg\,s$^{-1}$ (Wu et al. 2015\nocite{Wu15}). Numerous powerful QSOs have been detected outside the local Universe although their existence at high redshifts (e.g. Willott et al. 2010a\nocite{Willott10a}; 2010b\nocite{Willott10b}; De Rosa et al. 2011\nocite{DeRosa11}; Podigachoski et al. 2015\nocite{Podigachoski15}) poses challenges in our understanding of supermassive black hole growth in the early Universe (e.g. Volonteri 2012\nocite{Volonteri12}; Melia 2014\nocite{Melia14}). 

In recent years, much effort has been focused on studying the types of galaxies that powerful AGN reside in (e.g. Mainieri et al. 2011; Khandai et al. 2012\nocite{Khandai12}; Rawlings et al. 2013\nocite{Rawlings13}; Feruglio et al. 2014\nocite{Feruglio14}; Podigachoski et al. 2015\nocite{Podigachoski15}), with the ultimate aim to understand the origin of the black hole -- galaxy bulge mass relation in the nearby Universe (e.g. Magorrian et al. 1998\nocite{Magorrian98}; Ferrarese $\&$ Merritt 2000\nocite{FM00}), as well as the reason behind the co-evolution of the star-formation rate density and AGN accretion rate density (e.g. Boyle $\&$ Terlevich 1998\nocite{BT98}; Merloni et al. 2004\nocite{Merloni04}; Silverman et al. 2008\nocite{Silverman08}; Hirschmann et al. 2014\nocite{Hirschmann14}). Indeed, AGN hold a central role in models of galaxy formation and evolution, AGN feedback being essential in shaping the high mass end of the galaxy luminosity function (e.g. Croton et al. 2006, Bower et al. 2006).

\begin{figure*}
\epsfig{file=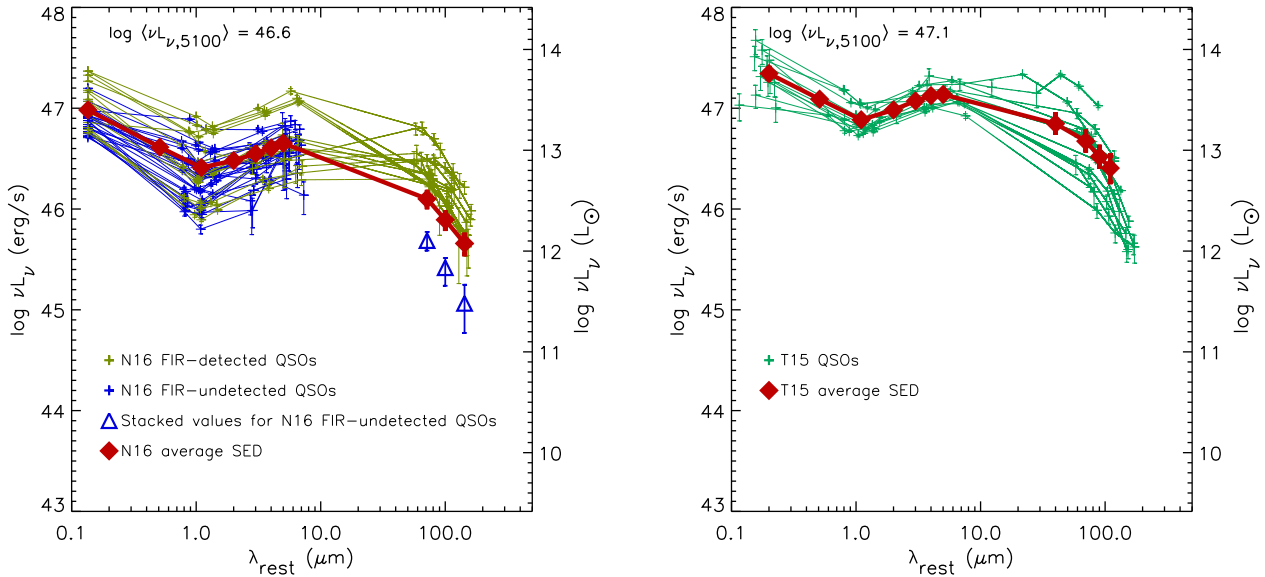,width=0.99\linewidth} 
\caption{SEDs for the Netzer et al. 2016 (left panel) and Tsai et al. 2015 (right panel) QSO samples. The red diamonds are the average SED of each sample. }
\label{fig:SEDs}
\end{figure*}

Since galaxy evolution is directly linked to the rate of star-formation and consumption of gas, as well as any processes, such as AGN feedback, which might interfere with it, one of the key properties of AGN host galaxies to chart is the star-formation rate (SFR). For unresolved sources, the AGN and stellar contributions to the global galaxy emission must be disentangled through spectral energy distribution (SED) fitting, a non-trivial task in its own right as it requires a-priori knowledge of the strength of either one component over the entire electromagnetic spectrum. In most cases, the host cannot be seen at X-ray to mid-IR wavelengths. For non Compton-thick AGN, X-ray emission is unequivocally AGN dominated as the X-ray luminosity from stellar processes, even for the most rapidly star-forming galaxies, is orders of magnitude lower than that of powerful AGN (e.g. Symeonidis et al. 2011\nocite{Symeonidis11b}; 2014\nocite{Symeonidis14b}). The optical spectra and optical broadband photometry of luminous unobscured QSOs are also AGN dominated, as is the case for the near and mid-IR, the origin of which is emission from the dusty torus (e.g.  Osterbrock $\&$ Ferland 2006; Rodriguez-Ardila $\&$ Mazzalay 2006). However, it is often assumed that the far-IR and submm part of the broadband SED, which in star-forming galaxies is a robust tracer of the SFR (e.g. Kennicutt 1998; Kennicutt $\&$ Evans 2012\nocite{KE12}), has but a minor contribution from the AGN and hence directly traces star-formation even in AGN hosts. Since the advent of the \textit{Herschel} space observatory which for the first time provided far-IR/submm data for many high redshift galaxies hosting powerful AGN, \textit{Herschel} far-IR SPIRE and/or PACS photometry has been frequently used to compute the SFRs of AGN hosts (e.g. Rovilos et al. 2012\nocite{Rovilos12}; Rosario et al. 2013\nocite{Rosario13}; Netzer et al. 2014\nocite{Netzer14}; 2016\nocite{Netzer16}; Drouart et al. 2014\nocite{Drouart14}; Delvecchio et al. 2015\nocite{Delvecchio15}; Tsai et al. 2015\nocite{Tsai15}; Mullaney et al. 2015\nocite{Mullaney15}).

Recently, in Symeonidis et al. (2016\nocite{Symeonidis16}; hereafter S16) we showed that the aforementioned assumptions regarding the AGN contribution to the far-IR/submm are not always valid. We derived an \textit{intrinsic} AGN SED from the optical to submm, by using mid-IR spectral lines to subtract the host galaxy component in a sample of powerful nearby QSOs, and noted that the intrinsic AGN SED retains a higher level of far-IR/submm power than previously thought. Our results indicated that for some galaxies hosting luminous ($\nu L_{\nu, 5100}$ or $L_{\rm X(2-10keV)}$ $\gtrsim$10$^{43.5}$ erg/s) AGN, the entire broadband SED could be AGN dominated at least up to 1000$\mu$m, the far-IR/submm emission being associated with AGN-heated dust at kpc scales.

In this letter I examine the SEDs of two samples consisting of the most luminous unobscured QSOs at $2<z<3.5$ in order to determine the origin of their far-IR/submm emission. The letter is laid out as follows: in section \ref{sec:sample}, I describe the samples and SED building and in section \ref{sec:results}, I report my results. The discussion and conclusions are presented in section \ref{sec:conclusions}. Throughout, I adopt a concordance cosmology of H$_0$=70\,km\,s$^{-1}$Mpc$^{-1}$, $\Omega_{\rm M}$=1-$\Omega_{\rm \Lambda}$=0.3.

\section{The sample}
\label{sec:sample}

I select two samples consisting of the most powerful, unobscured QSOs that have available mid and far-IR data, taken from Tsai et al. (2015; hereafter T15) and Netzer et al. (2016; hereafter N16); these QSOs are at $2<z<3.5$ and have rest-frame optical luminosity log\,$\nu L_{\nu}$(\rm 5100\,\AA) = 46.2--47.4. From N16, I chose their most luminous sub-sample of objects, which they define as log\,$\nu L_{\nu}$(\rm 1350\,\AA) $>$ 46.7, consisting of 15 far-IR (FIR) detected and 28 FIR-undetected QSOs. The T15 sample consists of 12 FIR-detected QSOs, on average about 0.5 dex more luminous in the optical than the N16 sample.

\begin{figure*}
\epsfig{file=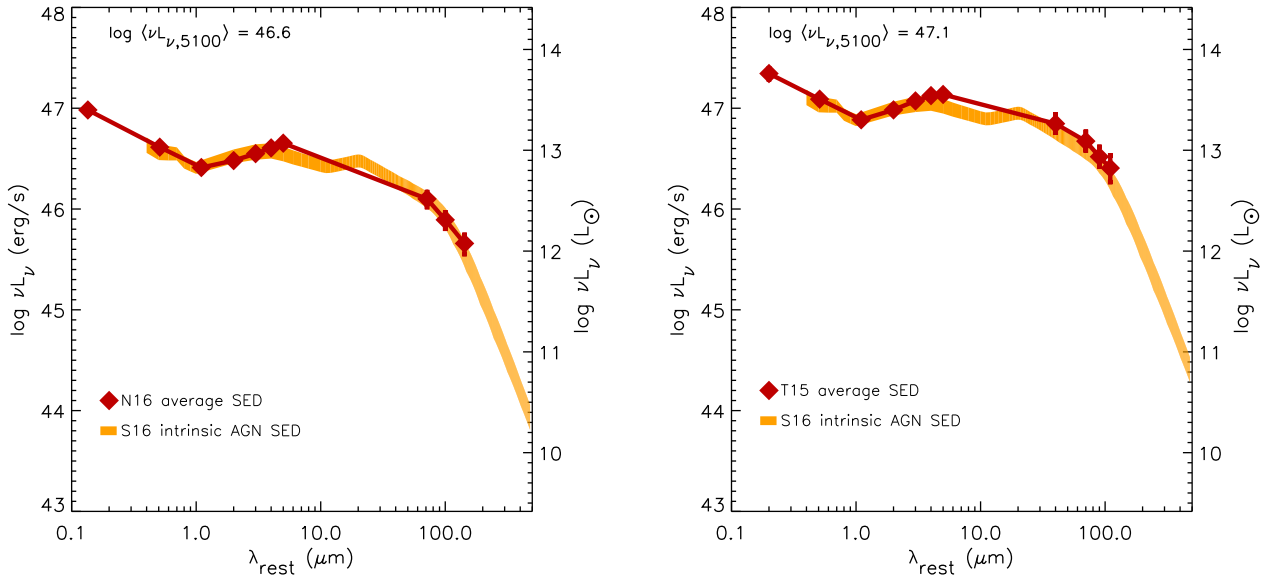,width=0.99\linewidth} 
\caption{The intrinsic AGN SED from Symeonidis et al. (2016; shaded orange region where the shading width represents the 1$\sigma$ bounds) normalised at 1$\mu$m to the average SEDs of the two samples (red diamonds). }
\label{fig:SEDs_comp}
\end{figure*}

Fig. \ref{fig:SEDs} shows their SEDs; for the N16 FIR-undetected QSOs, rest-frame stacked luminosities from N16 calculated at a nominal redshift of 2.5 are also plotted. To calculate an average SED for the whole N16 sample, I first compute the average SEDs of the FIR-detected and FIR-undetected sources separately. For the former, I logarithmically interpolate between the data points, obtaining rest-frame luminosities at 0.135, 0.51, 1, 2, 3, 4, 5, 71, 100, 143$\mu$m. The luminosities at each of those wavelengths are subsequently averaged. For the FIR-undetected sample, I logarithmically interpolate their SEDs up to 5$\mu$m, subsequently computing average rest-frame luminosities at 0.135, 0.51, 1, 2, 3, 4, 5$\mu$m and combining those with the FIR stacked luminosities at 71, 100, 143$\mu$m in order to get a full average SED for this subgroup. Once I have an average SED for each subgroup, I determine the weighted average SED of the whole N16 sample. To evaluate the errors on the average luminosities up to 5$\mu$m I combine the standard errors of the average luminosities of each sub-sample. For the FIR part of the SED, I combine the errors from stacking for the FIR-undetected subsample with the standard errors of the FIR-detected sub-sample. In both cases, I perform weighted error propagation, i.e. taking into account the number of sources in each subgroup. 

To obtain an average SED for the T15 sample (all of which are FIR-detected) I logarithmically interpolate between the data in order to compute rest-frame luminosities at 0.2, 0.51, 1, 2, 3, 4, 5, 40, 70, 90, 110$\mu$m and subsequently average those. Since I have full SEDs for all sources in the T15 sample, I calculate errors on the average luminosities by bootstrapping. The average SEDs are shown in Fig. \ref{fig:SEDs} as red diamonds. The mean 5100$\AA$ luminosity of the N16 and T15 samples is 10$^{46.6}$ erg/s and 10$^{47.1}$ erg/s respectively.

\section{Results}
\label{sec:results}

\subsection{SED comparison}
\label{sec:SEDcomparison}
To explore the origin of the broadband emission in these QSOs, I compare the average SED of each QSO sample to the intrinsic AGN SED derived in S16, normalising the latter at 1$\mu$m rest-frame (Fig. \ref{fig:SEDs_comp}). The S16 SED is the \textit{average}, \textit{intrinsic}, broadband emission from the AGN in a sample of optically luminous ($L_{\rm 5100}>$10$^{43.5}$\, erg/s), unobscured, radio-quiet, $z<0.18$, QSOs from the Palomar Green survey (we refer the reader to S16 for more details). As the S16 SED is an \textit{average}, it is more appropriate to compare it to the \textit{average} emission of uniformly selected samples of QSOs as opposed to individual SEDs. Fig. \ref{fig:SEDs_comp} shows this comparison. I find that the average SEDs of the two QSO samples examined here are not significantly different to the intrinsic AGN SED derived in S16; no datapoint is more than $1.3 \sigma$ away from the S16 SED. The maximum difference is at 5$\mu$m where the N16 and T15 average SEDs are 26 and 28 per cent higher than the S16 SED respectively, suggesting a stronger mid-IR bump, often seen in high luminosity AGN (e.g. Leipski et al. 2014\nocite{Leipski14}). Below 5$\mu$m the SEDs are offset by less than 16 per cent, and longwards of 5$\mu$m they are offset by less than 24 per cent. 

\begin{figure}
\epsfig{file=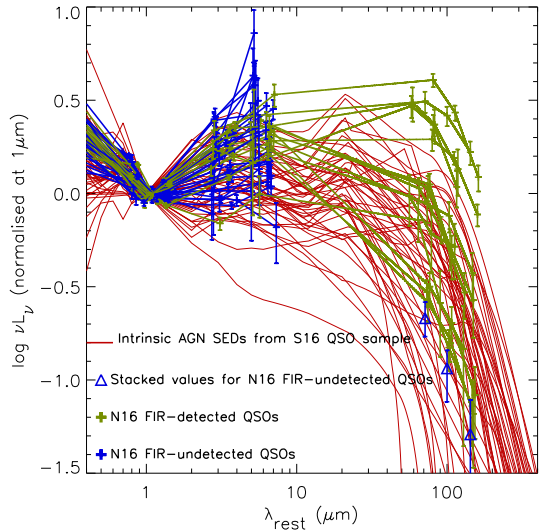 ,width=0.99\linewidth} 
\caption{The \textit{individual, intrinsic} AGN SEDs from S16 shown as red curves, compared with the individual N16 QSO SEDs (blue points for far-IR undetected sources and green points for far-IR detected sources)}
\label{fig:NetzerSEDs_comp}
\end{figure}
\begin{figure}
\epsfig{file=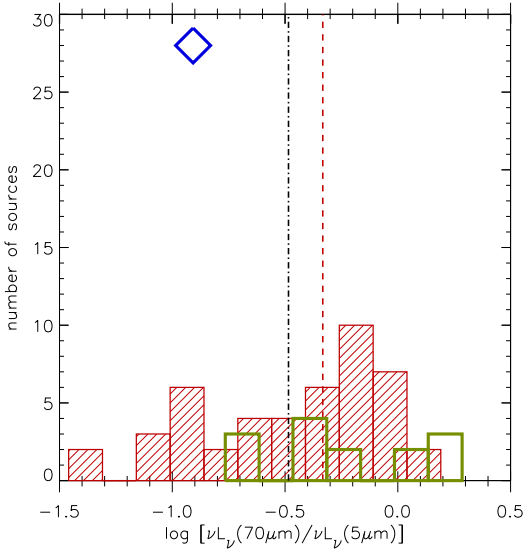 ,width=0.99\linewidth} 
\caption{Distribution of the log[$\nu L_{\nu(70\mu m)}$ to $\nu L_{\nu(5\mu m)}$] ratio for the S16 sample (red hatched histogram) compared to the N16 sample (blue point for the FIR-undetected sources using the stacked values and green histogram for the FIR-detected sources). The dashed red line shows the log of the average $\nu L_{\nu(70\mu m)}$/$\nu L_{\nu(5\mu m)}$ ratio for the S16 sample and the black dot-dashed line shows the log of the weighted average $\nu L_{\nu(70\mu m)}$/$\nu L_{\nu(5\mu m)}$ ratio for the N16 sample (including FIR-detected and FIR-undetected sources). }
\label{fig:NetzerSEDs_comp_hist}
\end{figure}
\begin{figure}
\epsfig{file=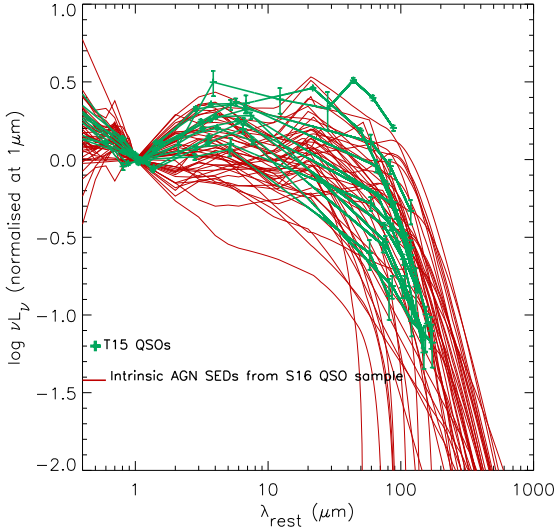 ,width=0.99\linewidth} 
\caption{The \textit{individual, intrinsic} AGN SEDs from S16 shown as red curves, compared with the individual T15 QSO SEDs (green points)}
\label{fig:TsaiSEDs_comp}
\end{figure}
\begin{figure}
\epsfig{file=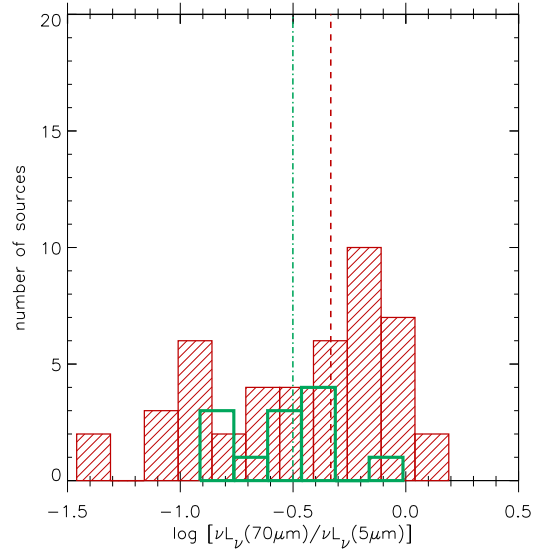 ,width=0.99\linewidth} 
\caption{Distribution of the log[$\nu L_{\nu(70\mu m)}$ to $\nu L_{\nu(5\mu m)}$] ratio for the S16 sample (red hatched histogram) compared to the T15 sample (green histogram). The log of the average $\nu L_{\nu(70\mu m)}$/$\nu L_{\nu(5\mu m)}$ ratio is shown with a red dashed line for the S16 sample and a green dot-dashed line for the T15 sample.}
\label{fig:TsaiSEDs_comp_hist}
\end{figure}

To develop this comparison further, I also examine the distribution of individual SEDs in the N16 and T15 samples against the distribution of the \textit{individual, intrinsic} (i.e. with the host component removed) AGN SEDs of the S16 sample (Figs \ref{fig:NetzerSEDs_comp} to \ref{fig:TsaiSEDs_comp_hist}). Fig. \ref{fig:NetzerSEDs_comp} shows that the majority of the N16 SEDs reside within the range of the S16 intrinsic AGN SEDs, except for 5 sources. Out of those 5 sources, some have higher 5$\mu$m luminosities, suggesting that the bright far-IR emission might be linked to the elevated mid-IR emission rather than an additional contribution from star-formation. To explore this in more detail, I plot the the distribution of far-to-mid-IR colour (70 to 5$\mu$m luminosity ratio; $\nu L_{\nu(70)}$/$\nu L_{\nu(5)}$) for the N16 sample, in comparison to that of the S16 sample (Fig. \ref{fig:NetzerSEDs_comp_hist}). Out of the 5 N16 sources with outlying SEDs, only 3 (7 per cent of the sample) have far-to-mid-IR colours which extend further than the tails of the distribution of the S16 sample.
It is possible that the different SED slopes of these 3 sources are due to an additional star-formation component. Removing these sources from the N16 sample, shifts the far-IR (70--150$\mu$m) luminosities of the average N16 SED to lower values by 11--15 per cent, nevertheless the N16 average SED remains consistent with the S16 SED; no datapoint from the N16 SED is more than $2 \sigma$ away from the S16 SED. 

I perform the same comparison for the T15 sample in Figs \ref{fig:TsaiSEDs_comp} and \ref{fig:TsaiSEDs_comp_hist}, finding no significant differences in the SEDs and distribution in far-to-mid-IR colour of the T15 sample when compared with the S16 sample.

The results presented here indicate that the AGN in the N16 and T15 QSOs could account for \textit{the entire} broadband emission from the UV to submm, if indeed the intrinsic AGN emission in these samples is consistent with the one derived in S16 for less powerful, lower redshift QSOs.

\begin{figure*}
\epsfig{file=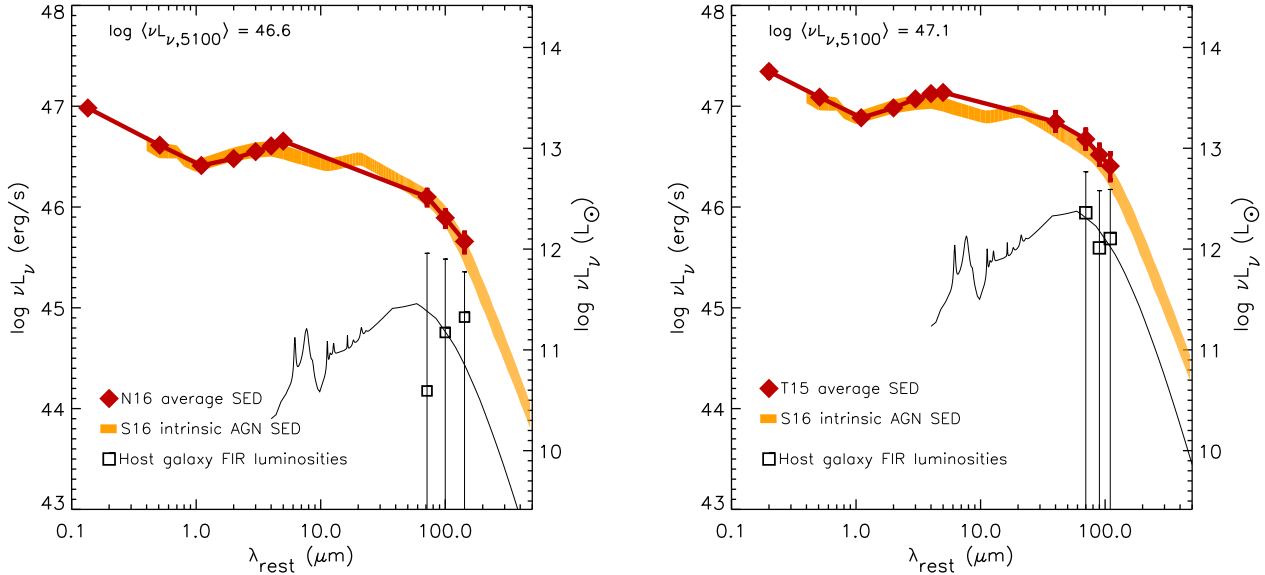,width=0.99\linewidth} 
\caption{The intrinsic AGN SED from Symeonidis et al. (2016; shaded orange region where the shading width represents the 1$\sigma$ bounds) normalised at 1$\mu$m to the average SEDs of the two samples (red diamonds). The S16 intrinsic AGN SED is subtracted from the average N16 SED at 71, 100, 143$\mu$m (left panel) and from the average QSO T15 SED at 70, 90, 110$\mu$m (right panel) in order to obtain the rest-frame FIR luminosities of the host galaxy SED (black open squares). An M82 SED template (from the GRASIL library; Silva et al. 1998) is also shown (black curve) fitted onto the rest-frame FIR host luminosities.}
\label{fig:SEDs_comp_plushost}
\end{figure*}

\subsection{How much star-formation is there?}
\label{sec:SF}
To investigate how much star-formation is present in the N16 and T15 QSOs, the S16 intrinsic AGN SED is subtracted from the average QSO T15 SED at 70, 90, 110$\mu$m and from the average N16 SED at 71, 100, 143$\mu$m, in order to obtain the rest-frame FIR luminosities of the host galaxy SED; the FIR host galaxy luminosities are shown in Fig. \ref{fig:SEDs_comp_plushost} fitted with an M82 SED template (from the GRASIL \footnote{http://adlibitum.oats.inaf.it/silva/grasil/modlib/modlib.html} library; Silva et al. 1998). I find that the total IR luminosity ($L_{\rm IR}$; 8-1000$\mu$m) attributed to the host accounts for 0--9 per cent (within $1\sigma$) of the $L_{\rm IR}$ of the N16 average SED and 0--20 per cent (within $1\sigma$) of the $L_{\rm IR}$ of the T15 average SED. This suggests that the AGN is responsible for the majority of the total infrared emission (over 80 per cent). Using the Kennicutt (1998) relation to convert $L_{\rm IR}$ to an SFR, I find that the upper limit (3$\sigma$) on the SFR for the T15 QSO sample is 3165\,M$_{\odot}$/yr and for the N16 sample it is 643\,M$_{\odot}$/yr. Although this indicates that the star-formation rate could be substantial, a contribution from star-formation is not required to account for the infrared emission.

\subsection{The origin of the far-IR emission in the most luminous QSOs}

\begin{figure*}
\epsfig{file=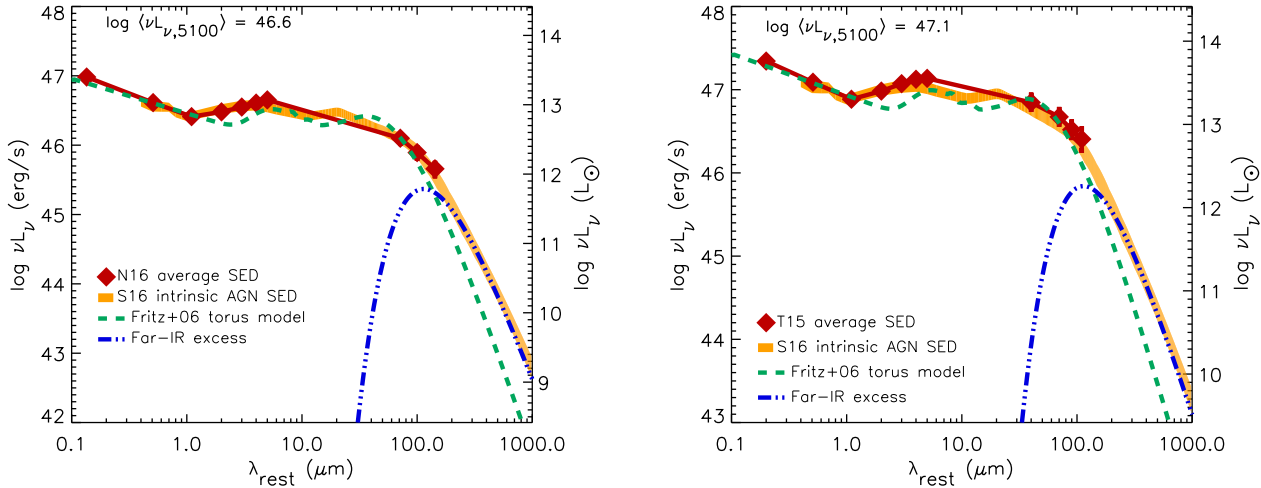,width=0.99\linewidth} 
\caption{The intrinsic AGN SED from Symeonidis et al. (2016; shaded orange region where the shading width represents the 1$\sigma$ bounds) normalised at 1$\mu$m to the average SEDs of the two samples (red diamonds). The best-fit torus model from the Fritz+06 library (green dashed curve) is subtracted from the intrinsic AGN SED in order to obtain the far-IR excess component (blue dot-dashed curve). The far-IR excess component corresponds to the infrared emission from dust at kpc scales heated by the AGN. }
\label{fig:SEDs_comp_plustorus}
\end{figure*}

In the previous sections (\ref{sec:SEDcomparison} and \ref{sec:SF}), I showed that a star-formation component is not necessary to explain the IR luminosity in the two samples of luminous QSOs studied here and that the emission in the infrared can be attributed to dust heated predominantly by the AGN. In the near/mid-IR, the emission is expected to come from dust in the torus, heated to near-sublimation temperatures. In the far-IR/submm, there is a contribution from the Rayleigh-Jeans tail of the torus dust emission, however, in S16 we argued that there is also an additional component of dust which is cooler than that found in the AGN torus, likely dust at kpc scales. We thereafter refer to this as the \textit{far-IR excess component} to describe infrared emission from dust not located in the AGN torus, but heated by the AGN. As also discussed in S16, this component, which is inherently included in the S16 intrinsic AGN SED, has been missing from previous empirical and modelled SEDs designed to represent the intrinsic AGN emission. 

To estimate the mass, size and energy of the far-IR excess component, one can subtract a torus model from the intrinsic AGN SED. The best model to chose would be one which represents a maximum contribution in the far-IR from the torus, in order to obtain lower limits on the mass, size and energy estimates of the far-IR excess component. In S16 we examined three widely used AGN torus model libraries (torus models from H{\"o}nig $\&$ Kishimoto 2010, and accretion disc+torus models from Fritz et al. 2006 and Siebenmorgen et al. 2015) fitting them to the intrinsic AGN SED; figure 13 in S16 shows the ones which are best fit to the intrinsic AGN SED. Here I chose to use the best fit torus model of Fritz et al. (2006; see figure 13 in S16 and Fig.\ref{fig:SEDs_comp_plustorus} this work) as it gives the highest level of far-IR emission for an unobscured, face-on AGN with optical power that matches that of the S16 intrinsic AGN SED. As stated above, this implies that the mass, size and energy estimates of the far-IR excess component (see below) should be thought of as lower limits. 

I subtract the Fritz et al. (2006) torus model from the S16 intrinsic AGN SED, normalised to the average SED of each QSO sample (Fig.\ref{fig:SEDs_comp_plustorus}). The subtraction is performed in the 75--1000$\mu$m region only, as at $\sim$75$\mu$m the torus SED crosses the intrinsic AGN SED. The far-IR excess component can be approximated by a greybody (emissivity $\beta$=1.5), which peaks at 114$\mu$m (in $\nu L_{\nu}$) corresponding to a dust temperature of 23\,K (the equivalent temperature for a blackbody with the same peak is 34\,K). The total amount of energy it carries is $2.6\times10^{45}$\,erg/s ($\pm 1\times10^{45}$) for the N16 sample and $7.7\times10^{45}$\,erg/s ($\pm 3\times10^{45}$) for the T15 sample, amounting to 80--87 per cent (within 1$\sigma$) of the total QSO power produced longwards of 200$\mu$m. This translates to 5--10 per cent (within 1$\sigma$) of the total luminosity of the QSO (integrated in the 0.1--1000$\mu$m range) being emitted in the far-IR. 

Its dust mass, $M_{\rm dust}$, is calculated as follows: 
\begin{equation}
M_{\rm dust}=\frac{f_{\nu, \rm rest} D_{\rm L}^2}{B(\nu_{\rm
    rest},T_{\rm dust, rest})
  \kappa_{\rm rest}} 
\end{equation}
where $D_{\rm L}$ is the luminosity distance, $B(\nu_{\rm
  rest},T_{\rm dust, rest})$ is the black body function (in units of
flux density), $f_{\nu,\rm rest}=\frac{f_{\nu,\rm obs}}{(1+z)}$, $\kappa_{\rm rest}=\kappa_{850\mu m} (\frac{\nu_{\rm rest}}{\nu_{\rm
    850\mu m}})^{\beta}$ and $\kappa_{850\mu m}$=0.0431\,m$^2$\,kg$^{-1}$  taken from Li $\&$ Draine (2001\nocite{LD01}). Here, I take $\nu_{\rm rest}$ as the rest-frame frequency equivalent to 850\,$\mu$m (observed), $\beta=1.5$ and $T_{\rm dust, rest}$=34\,K. $f_{\nu,\rm obs}$ at 850\,$\mu$m is computed from the SED of the far-IR excess component, using $z_{\rm mean}$=2.7 for the N16 QSOs and $z_{\rm mean}$=2.8 for the T15 QSOs. The dust mass calculated for the N16 sample is $1.1\times10^8$\,M$_{\odot}$ and for the T15 sample it is $3.2\times10^8$\,M$_{\odot}$. 

Finally, to estimate the extent of the dust region which corresponds to the far-IR excess emission, I use the Stefan-Boltzmann law for a greybody, $P=A\epsilon_{\rm \scriptscriptstyle GB} \sigma T^4$. In this case $P$ is the power carried by the far-IR excess component (as calculated above), $A$ is the area of the emitting greybody ($=4\pi r^2$, where $r$ is the radius), $\sigma$ is the Stefan-Boltzmann constant, $\epsilon_{\rm \scriptscriptstyle GB}$ is the emissivity of the greybody and $T$ is the temperature of the far-IR excess component. To compute $\epsilon_{\rm \scriptscriptstyle GB}$, I take the ratio of integrated greybody function (23\,K; $\beta=1.5$) to the integrated blackbody function (34\,K; normalised to the far-IR excess component peak), to be equal to the ratio of the emissivity of the greybody ($\epsilon_{\rm \scriptscriptstyle GB}$) to the emissivity of the blackbody ($\epsilon_{\rm BB}$=1); i.e. $\frac{GB}{\rm BB}=\epsilon_{\rm \scriptscriptstyle GB}$ (see also Greve et al. 2012\nocite{Greve12} for an equivalent method). I find $\epsilon_{\rm \scriptscriptstyle GB}=0.87$ and arrive at a radius, $r$, of 1.2\,kpc for the N16 sample and 2.2\,kpc for the T15 sample. Note that this is the minimum radius of the extended dust region heated by the AGN, assuming that the dust filling factor is 100 per cent. In practice the size of the dust region illuminated by the AGN will likely extend to several kpc, as the dust filling factor is much less than 100 per cent. 

The properties of the far-IR excess component in the two QSO samples are presented in table \ref{table:measurements}.

\begin{table}
\centering
\caption{The properties of the far-IR excess component for the QSO samples examined in this work. This component can be approximated by a greybody of peak wavelength 114$\mu$m (in $\nu L_{\nu}$), emissivity $\beta=1.5$ and dust temperature 23\,K. Its emission relates to dust at kpc scales heated by the AGN and makes up 5--10 per cent of the total AGN luminosity (integrated in the 0.1--1000$\mu$m range) and 80-87 per cent of the total power produced longwards of 200$\mu$m.  }
\begin{tabular}{|c|c|c|c|}
\hline 
QSO  & log\,$L_{\rm dust}$ & log\,$M_{\rm dust}$& $R_{\rm dust}$\\
sample& (erg/s) & (M$_{\odot}$)& (kpc) \\
\hline
N16 &45.4&8&$>$1.2\\
T16  &45.9&8.5&$>$2.2\\
\hline
\end{tabular}
\label{table:measurements}
\end{table}

\section{Discussion and Conclusions}
\label{sec:conclusions}

I find that, over the entire UV--submm wavelength range, the average SED shape of the most luminous unobscured QSOs at $2<z<3.5$ matches that of the average \textit{intrinsic} AGN SED derived in S16 from a set of lower redshift ($z<0.18$) QSOs, 2-3 orders of magnitude less luminous in the optical. Indeed it has been shown that the X-ray to mid-IR SEDs and spectra of QSOs do not evolve as a function of redshift or AGN luminosity (e.g. Wilkes et al. 1994\nocite{Wilkes94}; Mathur et al. 2002\nocite{Mathur02}; Grupe et al. 2006\nocite{Grupe06}; Hao et al. 2014\nocite{Hao14}), so one might expect that their far-IR SEDs are also unchanged. 

On the basis of the assumption that the intrinsic AGN emission as a function of wavelength is broadly independent of AGN luminosity (for $\nu L_{\nu, \rm 5100}$ or $L_{\rm X (2-10keV)}$\,$>$10$^{43.5}$ erg/s AGN) and redshift (at least up to $z\sim3.5$), my results indicate that, on average, the far-IR/submm emission in the most luminous QSOs at the tail of the $2<z<4$ optical luminosity function (Richards et al. 2006) is primarily produced by the AGN. This implies that for such systems, emission in the \textit{Herschel} bands and longer wavelengths such as those probed by SCUBA-2 at 850$\mu$m would be AGN dominated, thus wide-area \textit{Herschel} and SCUBA-2 surveys will contain a population of QSOs which are primarily powered by AGN. Moreover, the SFRs of such systems cannot be determined with X-ray to submm broadband photometry and other means such as infrared spectral lines will need to be used. Note that this does not mean that the host galaxies of luminous QSOs do not have star-formation, perhaps even a considerable amount: I calculate a 3$\sigma$ upper limit on the SFR of 3165\,M$_{\odot}$/yr and 643\,M$_{\odot}$/yr for the two QSO samples studied here. It simply means that the contribution from star-formation is not detected in the broadband SED as it is energetically less prominent than the AGN power at all wavelengths. This conclusion is in disagreement with the work of N16. N16 use an AGN torus template to decompose their QSO SEDs into an AGN torus component and a star-forming component. The torus template used by N16 has a  steeper far-IR slope than the S16 AGN SED, falling by about 0.5\,dex between 50 and 70$\mu$m, compared to a 0.1\,dex change in the S16 AGN SED. This implies that a smaller fraction of the far-IR emission is attributed to the AGN in the work of N16 and indeed they find that star-formation is a significant contributor to emission in the \textit{Herschel} bands. In S16, however, we argued that using torus templates in SED decomposition can significantly underestimate the AGN power in the far-IR, as they often lack an extended dust component, which would in turn boost the contribution required from star-formation. Other works which use SED decomposition to separate the AGN and star-formation contributions, also find that QSOs have infrared luminosities systematically larger than the IR luminosity produced by a torus component (e.g. Leipski et al. 2014\nocite{Leipski14}; Gruppioni et al. 2016\nocite{Gruppioni16}; Podigachoski et al. 2016\nocite{Podigachoski16}; Drouart et al. 2016\nocite{Drouart16}). Although traditionally this has been assigned to star-formation, here I argue that a significant fraction, and in some cases all, of this excess infrared luminosity can be attributed to AGN-heated dust at kpc scales.

Using a bolometric correction factor $f$=4, where $L_{\rm AGN, bol}= f \times \nu L_{\nu, 5100}$, and an efficiency of $\eta$=0.1 to convert AGN luminosity to black hole accretion rate (BHAR) as in N16, I calculate a 3$\sigma$ upper limit on the SFR/BHAR ratio of 23 for the N16 sample and 36 for the T15 sample. However, note that higher values of $f$ have also been reported ($f=5-13$; e.g. Elvis et al. 1994\nocite{Elvis94}; Kaspi, Brandt $\&$ Schneider 2000\nocite{Kaspi00}; Netzer 2003\nocite{Netzer03}; Marconi et al. 2004\nocite{Marconi04}; Netzer $\&$ Trakhtenbrot 2007\nocite{NT07}) suggesting that the SFR/BHAR upper limit could be even lower. A value of SFR/BHAR$\sim$500 corresponds to the black hole-galaxy bulge mass relation in the local Universe, i.e. black holes growing in tandem with their hosts, and SFR/BHAR=142 corresponds to $L_{\rm SF} \sim L_{\rm AGN}$ (e.g. Alexander $\&$ Hickox 2012\nocite{AH12}; N16). 
N16 report larger values of SFR/BHAR than the upper limit I calculate here, particularly for their FIR-detected QSOs, as they assume a greater level of contribution from star-formation in the far-IR. They also find that some sources in their sample are very close to the $L_{\rm SF} \sim L_{\rm AGN}$ regime. In contrast, my findings suggest that the most luminous QSOs at $2<z<4$ have $L_{\rm AGN}$ considerably larger than $L_{\rm SF}$ and are hence AGN-dominated. Using hydrodynamical simulations of galaxy mergers, Volonteri et al. (2015\nocite{Volonteri15}) propose that such AGN-dominated galaxies are likely to be either in the merger or remnant phase, with a BHAR at a given SFR about 20 times higher than for SF-dominated galaxies.

I find that at least 5-10 per cent of the bolometric AGN luminosity is absorbed by dust beyond the torus, at kpc scales, and re-emitted in the far-IR/submm, accounting for the major fraction of the global emission longwards of 200$\mu$m. This is consistent with Baron et al. (2016) who find that the typical optical-UV continuum slopes of optically unobscured QSOs are reddened by dust along the line of sight, corresponding to a total luminosity emitted in the FIR of the order of 15$\%$ of the AGN bolometric luminosity. I calculate that the extent of the dust region illuminated by the AGN in the most powerful QSOs is at least a few kpc, comparable to those reported for submm galaxies and other high redshift QSOs --- studies with the Plateau de Bure Interferometer, the SMA and ALMA (e.g. Riechers et al. 2009\nocite{Riechers09}; Bothwell et al. 2010\nocite{Bothwell10}; Younger et al. 2010\nocite{Younger10}; Carilli et al. 2010\nocite{Carilli10}; Hodge et al. 2013\nocite{Hodge13}; Wang et al. 2013\nocite{Wang13}; Simpson et al. 2015\nocite{Simpson15}) measure the dust FWHM to be 1--10\,kpc, with a mean/median within the lower half of this range. Moreover, the implied dust masses of the order of 10$^8$\,M$_{\odot}$ are comparable to those of star-forming IR-luminous galaxies (e.g. Santini et al. 2010; 2014; Dunne et al. 2011) and consistent with those calculated for other high redshift QSO samples (e.g. Bertoldi et al. 2003\nocite{Bertoldi03}; Beelen et al. 2006\nocite{Beelen06}; Wang et al. 2008\nocite{Wang08}; 2011\nocite{Wang11}). 

I find that the relative power carried by the far-IR/submm component is independent of AGN luminosity (for $\nu L_{\nu, \rm 5100}$ or $L_{\rm X (2-10keV)}$\,$>$10$^{43.5}$ erg/s AGN), implying a constant far-IR dust temperature. According to the Stefan-Boltzmann law, this suggests that the radius of the corresponding dust region must scale with $\sqrt L_{\rm AGN}$. For this dust to be directly heated by the AGN, it is likely to be within the ionisation cone of the AGN; indeed there is evidence for the presence of dust in the narrow line regions (NLRs) of AGN from spectropolarimetry data (e.g. Goodrich et al. 1989\nocite{Goodrich89}; 1992\nocite{Goodrich92}, Vernet et al. 2001\nocite{Vernet01}; Smith et al. 2004\nocite{Smith04}). Furthermore, my estimates for the size of the dust region and its dependence on the AGN luminosity are consistent with reported values of $<$20\,kpc for the extent of the NLR (e.g. Liu et al. 2013\nocite{Liu13}; Hainline et al. 2013\nocite{Hainline13}; 2014\nocite{Hainline14}) and its suggested scaling with $\sqrt L_{\rm AGN}$ (e.g. Bennert et al. 2002\nocite{Bennert02}; Hainline et al. 2013).

\bibliographystyle{mn2e}
\bibliography{references}

\end{document}